\pdfoutput=1

\documentclass[11pt]{article}

\usepackage[]{acl}

\usepackage{times}
\usepackage{latexsym}
\usepackage{enumitem}
\usepackage{graphicx}

\usepackage[T1]{fontenc}
\usepackage{CJKutf8}
\newcommand{\zh}[1]{\begin{CJK}{UTF8}{gbsn}#1\end{CJK}}
\usepackage[utf8]{inputenc}
\usepackage{microtype}

\title{A Weibo Dataset for the 2022 Russo-Ukrainian Crisis}

\author{Yi R. Fung, Heng Ji \\
  University of Illinois at Urbana-Champaign \\
  \texttt{\{yifung2,hengji\}@illinois.edu}}

\begin{document}
\maketitle
\begin{abstract}
Online social networks such as Twitter and Weibo play an important role in how people stay informed and exchange reactions. Each crisis encompasses a new opportunity to study the portability of models for various tasks ({\em e.g.}, information extraction, complex event understanding, misinformation detection, etc.), due to differences in domain, entities, and event types.
We present the \textbf{\underline{R}ussia-\underline{U}kraine Crisis \underline{W}eibo (RUW) dataset}, with over 3.5M user posts and comments in the first release\footnote{Our data is available at \url{https://github.com/yrf1/RussiaUkraine_weibo_dataset}}.
\end{abstract}

\section{Introduction}
The Russo-Ukrainian conflict stemmed from complex issues that added up since the disintegration of the Soviet Union era, including Ukraine's geo-political divide between its internal East and West \cite{karacsonyi2014east}, NATO eastward expansion \cite{sarotte2014broken}, and Russian interference with Ukraine's eastern breakaway regions \cite{kofman2017lessons}. The conflict escalated on February 24, 2022, with Russian forces entering Ukraine territory under the campaign of a special military operation \cite{Guardian_2022}. At the time of writing, the conflict is, unfortunately, still ongoing. People across the world have been following the news and sharing opinions regarding the situation online. 

The study of information discourse and propagation in online social networks is of particular interest to the research community \cite{sacco2015using}. Deeper analysis from information extraction \cite{wen2021resin}, argumentation mining \cite{lawrence2020argument,reddy2021newsclaims}, opinion mining \cite{ravi2015survey}, propagation pattern monitoring \cite{shu2020hierarchical}, and misinformation detection \cite{shu2019defend,fung2021infosurgeon}, etc., can help society better understand and manage crises events. To faciliate timely analysis, we thus publish an open dataset from Weibo posts relevant to the ongoing 2022 Russo-Ukrainian crisis, with continued daily updates. 

\section{Related Work}
Various previous attempts have explored online social networks, in the aspects of misinformation detection \cite{islam2020deep}, propaganda identification \cite{khanday2021identifying}, and general network characteristics during crises \cite{kim2018social}. However, news media and news dissemination tend to exhibit different properties under different time and places, leading to a diachronic bias, which limits the portability of existing models \cite{murayama2021mitigation}. While there exists a recent Russo-Ukrainian Crisis Twitter dataset \cite{2203.02955}, no such dataset exists yet for Weibo, the second largest news-oriented social media platform in the world. It is generally interesting to uncover patterns universal across platforms, regulated by a separate set of rules and participated by a separate set of users with different user behavior. Note that 
Twitter data (multilingual) involves English as the primary language, and Weibo data involves Chinese as the primary language.

\section{Dataset Collection}
We use the publicly available \texttt {weibo-scraper}\footnote{\url{https://pypi.org/project/weibo-scraper/}} library package to collect Weibo data on the Russian-Ukraine crisis, including the following information for each public user post. 

\begin{enumerate}[label=\roman*]
  \itemsep0em 
  \item \underline{User information}: profile ID, profile name, profile image, and profile description..
  \item \underline{Post content}: the main text, as well as any images and videos attached, if available.
  \item \underline{Post metadata}: the \# of likes and shares on the post.
  \item \underline{Post comments}: information similar to i-iii for all comments underneath a user post.
\end{enumerate}

\noindent Note, the weibo-scraper library supports querying Weibo tweets by user profile. Hence, we manually prepared a list of the top $120$ public user profiles, who have actively posted about and ranked amongst the top posts of trending hashtags related to the Russian-Ukraine crisis. The average number of reactions and shares from the posts of these users are $4338$ and $163$, respectively. 
Then, we performed a one-hop expansion to include all the users who commented underneath the post of these active profiles as well.

Since the crisis is still ongoing, the list of user profiles may likely change over the time, with inclusion of new active user profiles according to the situation. We initiated the data collection process on the final week of Feb 2022. As the development of events preceding a conflict breakout is also of interest, we retrospectively collected data within the last three months prior to the onset of the conflict as well. We will continue the data collection and update the data repository accordingly.

\section{Dataset Summary}
By March 8th, we have collected over $27,341$ Weibo posts, and over $3.5$ million corresponding Weibo comments from $107,797$ unique users, containing keywords relevant to the Russian-Ukraine crisis. In Table \ref{tab:keyword_distribution}, we show the Weibo post distribution based on the keywords in this Russia-Ukraine Crisis Weibo (RUW) dataset. Most of the Weibo posts contain the \zh{乌克兰} (Ukraine) and \zh{俄罗斯} (Russia) reference, followed by \zh{俄乌} (Russia-Ukraine). The two countries' leaders, \zh{普京} (Putin) and \zh{泽连斯基} (Zelensky), are mentioned less.
\begin{table}[hbt!]
    \centering
    \begin{tabular}{ |c | c| }
    \hline
     \textbf{Keyword} & \textbf{\# Weibo Posts} \\ 
     \hline
     \zh{乌克兰} (Ukraine) & 120436 \\
     \zh{俄罗斯} (Russia) & 95845 \\  
     \zh{俄乌} (Russia-Ukraine) & 53164 \\
     \zh{基辅} (Kiev) & 28748 \\
     \zh{普京} (Putin) & 21005 \\
     \zh{泽连斯基} (Zelensky) & 18109 \\
     \zh{俄方} (Russian side) & 10612 \\
     \zh{北约} (NATO) & 10025 \\
     \zh{乌方} (Ukrainian side) & 7181 \\
     \hline
    \end{tabular}
    \caption{Number of Weibo posts for each keyword, in descending order.}
    \label{tab:keyword_distribution}
\end{table}

\begin{figure}[h!]
    \centering
    \includegraphics[trim={1.8cm 2.4cm 0 2.9cm},clip,scale=0.6]{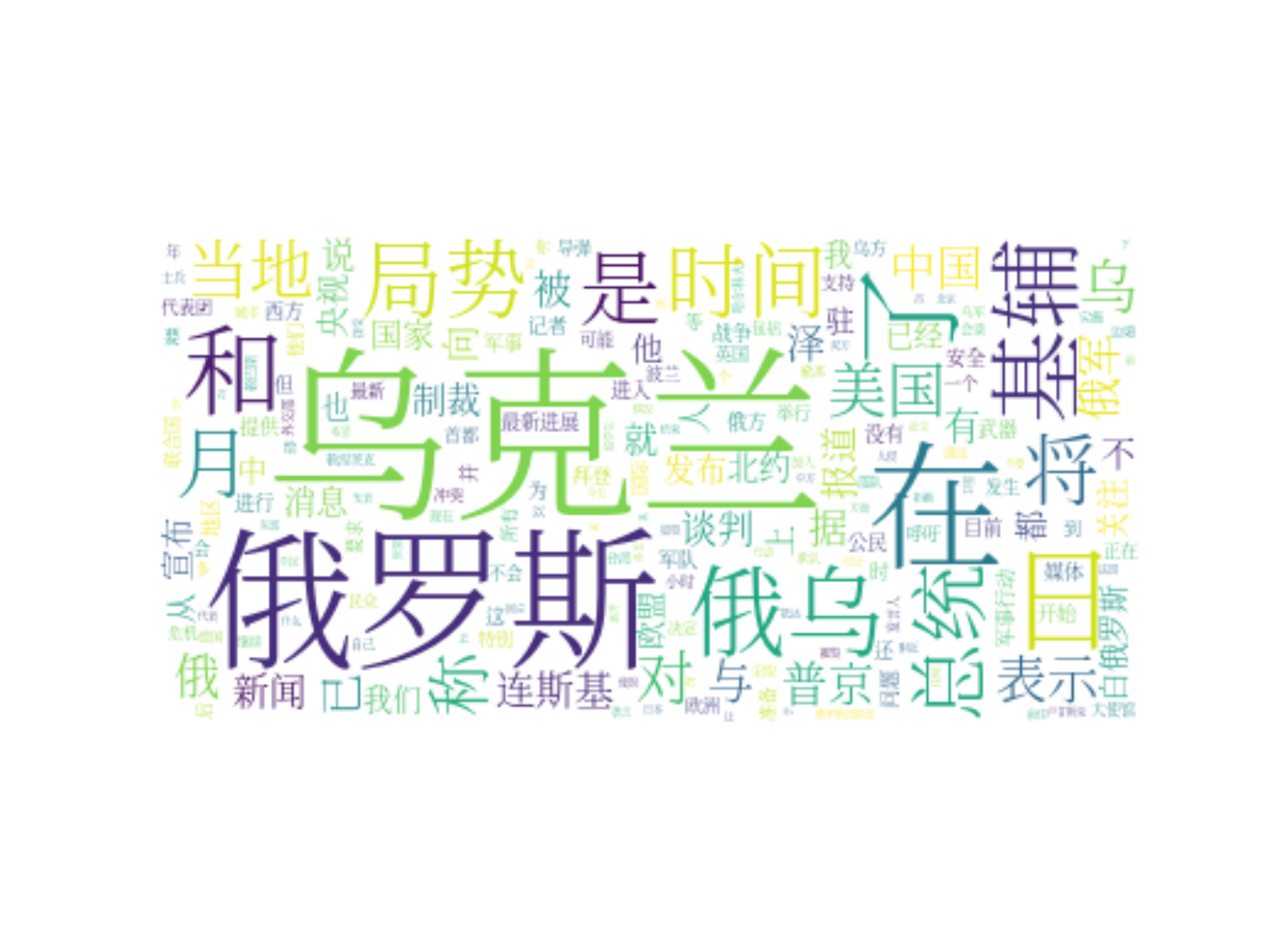}
    \caption{Word cloud of all Russo-Ukranian Weibo posts.}
    \label{fig:my_label}
\end{figure}

\noindent In addition, we show a word cloud of text from the Weibo posts in Figure \ref{fig:my_label}. Besides the named entities ({\em i.e.}, country name and person name), the prominence of words such as '\zh{局势} (situation)' and "\zh{制裁} (sanction)' suggests that most of tweets discuss the latest updates on the Russo-Ukrainian crisis.

\section{Dataset Usage}
There are many interesting task settings that can be experimented with, using our Russo-Ukranian Crises Weibo dataset. These include event clustering, false rumor detection, and evaluating the portability of news analytic methodologies across Twitter and Weibo domains. 

\section{Ethical Considerations}
We recognize that societal response in crises events may typically be polarized or contentious across different geo-political and socio-cultural regions. We urge researchers and analysts to be as objective as possible in their studies of the matter. 

\bibliography{references}



\end{document}